# Graph database while computationally efficient filters out quickly the ESG integrated equities in investment management

**Author 1 - Partha Sen**, Professor of Practice, GIBS Bangalore
+919002228212, research@parthasen.net

**Author 2 - Sumana Sen**, Founder, ESGlytics India, Bangalore
+ 919972632062, ssenfin@gmail.com

**Abstract**

*Design/methodology/approach*

This research evaluated the databases of SQL, No-SQL and graph databases to compare and contrast efficiency and performance. To perform this experiment the data were collected from multiple sources including stock price and financial news. Python is used as an interface to connect and query databases (to create database structures according to the feed file structure, to load data into tables, objects, to read data , to connect PostgreSQL, ElasticSearch, Neo4j.

*Purpose*

Modern applications of LLM (Large language model) including RAG (Retrieval Augmented Generation) with Machine Learning, deep learning, NLP (natural language processing) or Decision Analytics are computationally expensive. Finding a better option to consume less resources and time to get the result.

*Findings*

The Graph database of ESG (Environmental, Social and Governance) is comparatively better and can be considered for extended analytics to integrate ESG in business and investment.

*Practical implications*

A graph ML with a RAG architecture model can be introduced as a new framework with less computationally expensive LLM application in the equity filtering process for portfolio management.

*Originality/value*

Filtering out selective stocks out of two thousand or more listed companies in any stock exchange for active investment, consuming less resource consumption especially memory and energy to integrate artificial intelligence and ESG in business and investment.

**Keywords**

Database, ESG(Environmental, Social and Governance), Graph ML, RAG, AI architecture

**Introduction**

ESG integration represents a holistic approach to investment analysis and portfolio management, embedding environmental, social, and governance factors at every step of the investment process. The ESG integration framework (Appendix 1) suggests a multi-layered methodology where the central tenet of ESG integration informs key processes and detailed components, which are then operationalized through specific techniques. At the foundational level, research, asset allocation, risk management, and portfolio construction are identified as the cardinal processes that are influenced by ESG integration. Research activities, including the development of a materiality framework and the use of centralized research dashboards, are crucial for identifying ESG factors that are pertinent to the investment's value proposition. Asset allocation decisions, both strategic and tactical, are then adjusted to incorporate ESG insights, reflecting a dual focus on immediate responsiveness and long-term alignment with ESG goals. Risk management processes, underscored by the assessment of financial and ESG risk exposures, emphasize the importance of understanding and mitigating potential ESG risks. Portfolio construction, the culminating process, integrates ESG considerations into the selection and weighting of investments, ensuring the portfolio reflects the targeted ESG profile (CFA Institute and PRI, 2018)

Usually the portfolio is optimized on the financial performance but this study extends by creating a database to track the ESG and Economics related news along with financial data (Jain, 1988). News used for tracking this experiment was conducted using own mined news from multiple sources. The purpose of doing this research is to help to decide on how to store these data efficiently creating better relationships and to read complex queries effectively with better performance (appendix 2).

**Literature review**

The efficient markets hypothesis states that stock prices reflect all known information, including economic news. A linear or nonlinear model where coefficients are not zero, then this would provide evidence against the efficient markets hypothesis. The model can also be used to assess the speed at which stock prices adjust to new economic information. One term of the model can be used to measure the impact of economic news that has already been released (Pearce & Roley, 1985). One of the main reasons why media can lead to increased volatility is that it can amplify the spread of information. When investors see that other investors are buying or selling a particular stock, they may be more likely to do the same, even if they do not have any independent information about the stock. In some cases,the media can actually help to reduce volatility by providing investors with access to more information and by allowing them to communicate with each other more easily. However, the overall impact of the media on stock market volatility is still a matter of debate (Dahal, Pokhrel & Gaire et. al, 2023). Some studies have found that the market reacts significantly to firm-specific corporate news, while others have found that the reaction is small or even insignificant. This suggests that the market reaction to firm-specific corporate news may vary depending on the type of news, the company involved, and the market conditions (Dogra et. al, 2022).

PostgreSQL is a powerful open-source relational database management system (RDBMS) known for its advanced features, extensibility, and stability. PostgreSQL follows a traditional client-server model. The clients (applications or users) communicate with the PostgreSQL server to access and manipulate data. The PostgreSQL server is the core component responsible for managing data storage, retrieval, and processing having a database cluster within which multiple databases can be created. Each database operates independently and contains its schema, tables, views, functions, and data (Thomas, 2014). Elasticsearch is a distributed search engine that can scale over clusters of nodes. Text analysis in ElasticSearch is a crucial process that occurs before text data is stored and indexed. It involves breaking down the text into smaller units called tokens, which are then stored in data structures to optimize the searching process (Alberto, 2022). By default, ElasticSearch uses the standard analyzer, which employs a Standard Tokenizer and a lowercase token filter to make searches case-insensitive. The Query API is a significant part of the ElasticSearch API, and the query process can be divided into two main phases: the scatter phase and the gather phase (Kathare, Nikita, Vinati et. al., 2020).

Neo4j is a graph database management system that is used for building intelligent platforms with native graph performance. It allows users to find patterns and hidden connections in their data, navigate large hierarchies and multi-level data, and more. Neo4j is a native graph database, which means that it implements a true graph model all the way down to the storage level (Neo4j, n.d.)

**Research Methodology**

This research is conducted using Python as an interfacing source to connect and query databases (to create database structures according to the feed file structure, to load data into tables,objects, to read data for complex query outputs, to connect Python to all three databases [PostgreSQL, ElasticSearch, Neo4j] and compare steps for working with all the mentioned DBs in terms of performance, complexity and data storage size.

Loading of two months stock price data for different stocks done by reading multiple CSV files and combining them in python. Then SQL insert statements written to load the csv in PostgreSQL. Then csv was transformed to dictionary/json format to load in ElasticSearch.

Loading the two months of news articles extracted in python using googlenews package from various news articles with a format of portal name, date and content from the file and put that in the dataframe. In postgresql a '*tsvector*' column was created for fulltext search then dataframe to dictionary/json format was transformed to load in ElasticSearch. Cipher codes were written to load data into the Neo4j database. In Neo4j a fulltext search index for searching the text was created. ElasticSearch was based on-premise local system so exploration was restricted to a single node (Appendix 5 - Query Process).

**Results**

Neo4j is performing significantly better (Appendix 3) in terms of response time (measured in milliseconds) for News Search. It has the lowest values across all three metrics, indicating that it retrieves results faster than PostgreSQL and ElasticSearch for these operations. Specifically, for News Search, Neo4j is showing a response time of 8.45 ms which is much lower compared to 157.52 ms for PostgreSQL and 1260 ms for ElasticSearch. Similarly, for the News & OHLC relation, Neo4j's response time is 5.76 ms, which is again lower than that of the other two databases. Lastly, for ESG mentioned stocks, Neo4j has a response time of 23 ms, which is significantly lower than PostgreSQL's 15 ms and much lower than Elasticsearch 295 ms. This suggests that Neo4j, which is a graph database, is optimized for these kinds of searches and relations, likely due to its ability to quickly traverse nodes and relationships in the graph. The nature of the operations mentioned—searching news, relating news to OHLC (Open, High, Low, Close) data, and identifying stocks mentioned in an ESG (Environmental, Social, and Governance) context—seem to be well-suited to the strengths of a graph database. Graph databases like Neo4j are designed for rapid traversal of connected data, which can be a significant advantage for the mentioned queries, where relationships between pieces of data are key.

Neo4j has the lowest maximum CPU usage at 51% and an average CPU usage of 20%, which is significantly less than PostgreSQL and ElasticSearch (Appendix 4). Lower CPU usage can imply that Neo4j is more efficient in its operations, potentially leading to better performance in terms of response times. Neo4j's lower CPU usage suggests that it may be more efficient at processing operations, which can lead to faster response times—this correlates with the performance data. Neo4j's lower CPU usage could contribute to better time performance compared to PostgreSQL and ElasticSearch, even though it tends to use more memory on average.

**Discussion**

Graph databases like Neo4j excel at managing and querying highly connected data, making them particularly suitable for applications that involve complex relationships and pattern recognition. Graph Machine Learning (Graph ML) leverages these strengths to apply machine learning algorithms to graph-structured data. Graph databases can be used to efficiently extract features that are relevant for machine learning models. This includes node-level features (such as the properties of individual entities), edge-level features (such as the strength and characteristics of relationships), and higher-order features derived from the graph structure (such as centrality measures). Graph databases can be used to prepare data by defining subgraphs or extracting adjacency matrices, which are common inputs for Graph ML models (Hugging Face, 2023). Graph ML often involves converting graph data into a low-dimensional space (embedding) while preserving its topological structure. Graph databases can support this process by enabling the efficient computation of embeddings using techniques like node2vec, graph convolutional networks (GCNs) (Ahmedt-Aristizabal et al., 2021).

*Graph Machine Learning (Graph ML) into a RAG architecture* can be enhanced by either the retrieval component, the generation component, or both graph-based data and algorithms.

Graph-Based Retrieval is improved because of this Graph database (OpenAI, 2023). Also the graph embeddings can represent the knowledge in a vector space, which can be searched using vector similarity measures. Also graph algorithms can prioritize or filter the information retrieved. Other improvements like Augmented Generation with Graph ML by graph structured prompts and knowledge graph in decoding. Training process can be improved by graph regularized training and joint learning of retriever and generator through GCNs. Both loop and scaling criteria are expected to be improved.

**Conclusion**

Graph databases like Neo4j optimize resource usage and enhance performance in tasks like ESG-integrated Equity Filtering, crucial in investment management. Their ability to efficiently manage and query interconnected data enables faster, more accurate analyses, making them indispensable tools for sophisticated financial decision-making strategies. Neo4j's superior performance in terms of response time and CPU usage underscores its optimization for handling complex, interconnected data. Its efficiency in graph operations enhances applications that rely on deep relationship analysis, like those using Graph Machine Learning. When integrated into RAG architectures, Neo4j's capabilities in graph-based retrieval and generation can significantly improve the system's overall functionality, making it a powerful tool for ESG-integrated Equity Filtering.

# References

Ahmedt-Aristizabal, D., Armin, M. A., Denman, S., Fookes, C., & Petersson, L. (2021). A Survey on Graph-Based Deep Learning for Computational Histopathology. *arXiv.* https://arxiv.org/pdf/2107.00272.pdf

CFA Institute and PRI. (2018). Guidance and Case Studies for ESG Integration: Equities and Fixed Income. Retrieved from https://www.unpri.org/investor-tools/guidance-and-case-studies-for-esg-integration-equities-and-fixed-income/3622.article.

Dahal, K. R., Pokhrel, N. R., Gaire, S., Mahatara, S., Joshi, R. P., Gupta, A., Banjade, H. R., & Joshi, J. (2023). A comparative study on the effect of news sentiment on stock price prediction with deep learning architecture. *PLOS ONE, 18(4)*, e0284695. https://doi.org/10.1371/journal.pone.0284695.

Dogra, V., Alharithi, F. S., Álvarez, R. M., Singh, A., & Qahtani, A. M. (2022). NLP-Based application for analyzing private and public banks stocks reaction to news events in the Indian Stock Exchange. *Systems, 10(6)*, 233. https://doi.org/10.3390/systems10060233.

Jain, P. C. (1988). Response of hourly stock prices and trading volume to economic news. *Journal of Business,* 219-231.

Kathare, N., Reddy, O. V., & Prabhu, V. (2020). A comprehensive study of elasticsearch. *International Journal of Science and Research (IJSR)*.

Neo4j. (n.d.). Note on native graph databases. Retrieved from https://neo4j.com/blog/note-native-graph-databases/

OpenAI. (2023, December 8). RAG with a Graph database. *OpenAI Cookbook*. Retrieved from https://cookbook.openai.com/examples/rag_with_graph_db

Pearce, D. K., & Roley, V. V. (1985). The response of stock prices to economic news. *The Journal of Business, 58(1)*, 49-67. https://www.jstor.org/stable/2352909.

Thomas, S. M. (2014). *PostgreSQL 9 High Availability Cookbook*. O'Reilly. https://learning.oreilly.com/library/view/postgresql-9-high/9781849516969/?ar.

Hugging Face. (2023, January 3). Introduction to Graph Machine Learning. Hugging Face Blog. https://huggingface.co/blog/intro-graphml

# Appendix
Appendix 1 - ESG integration framework

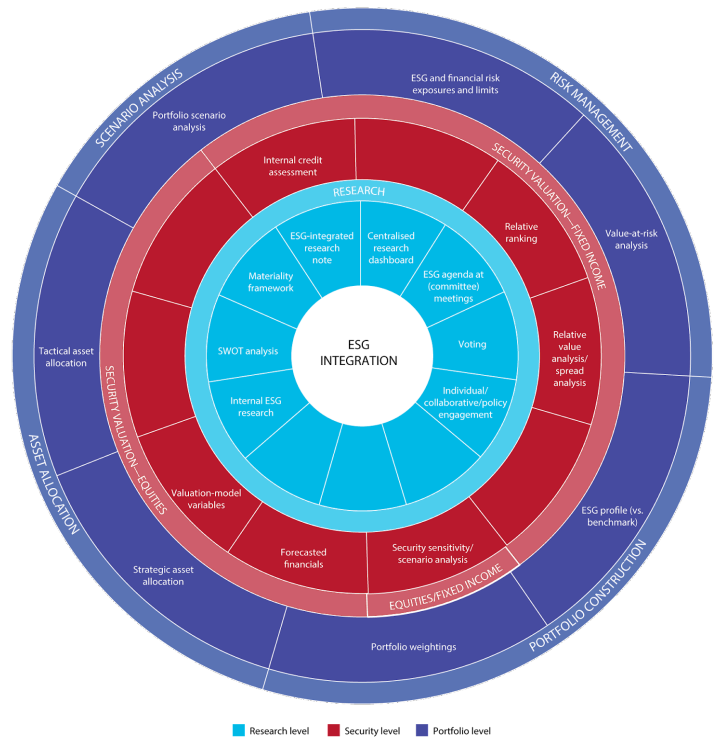

Appendix 2 - Five testing queries for Benchmarking
1. Which news articles have the required ESG list and for which stock names and when?
2. How this news affected the OHLC price for the entire news article day and should return all time series for that day?
3. How are the remaining unaffected stocks for the same day and return all time series for that day?
4. How will the Q2 & Q3 for next 5 days fetch the OHLC results?
5. How ESG Matched Stocks related to other stocks in the same industry and fetch the OHLC results of those stocks?

Appendix 3 - Time comparison

Appendix 4 - Memory and CPU usage comparison

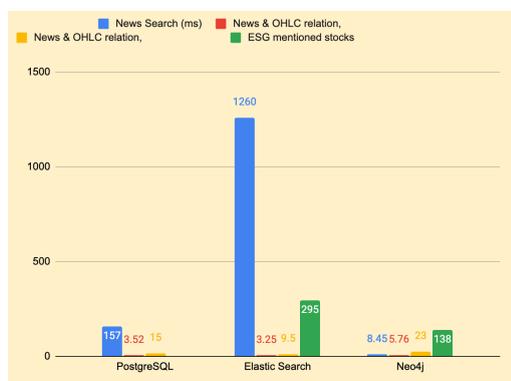

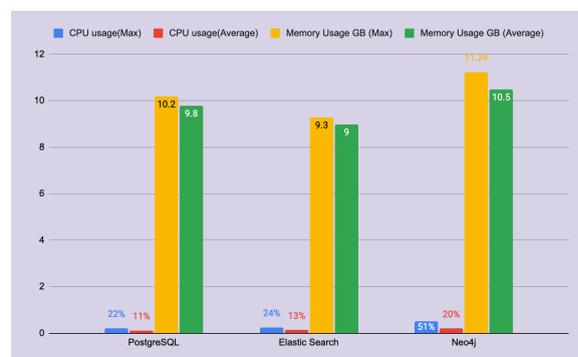

# Appendix 5 - Query Process

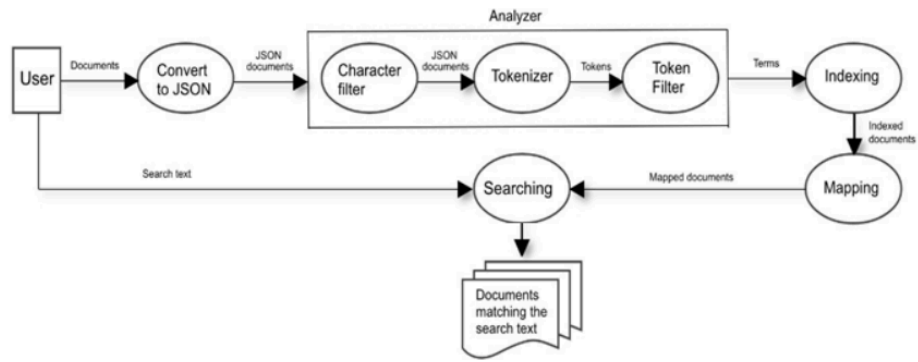

(Kathare, Nikita, O. Vinati Reddy, and Vishalakshi Prabhu 2020)

Elastic Search result	PostgreSQL result	Neo4j result

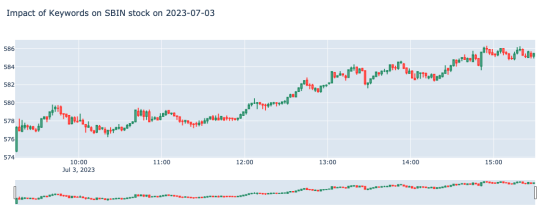

## SBIN

## TATASTEEL

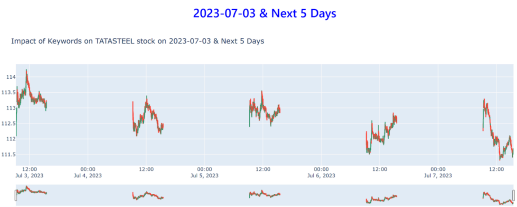

## Financial Sector

## Metals and Mining

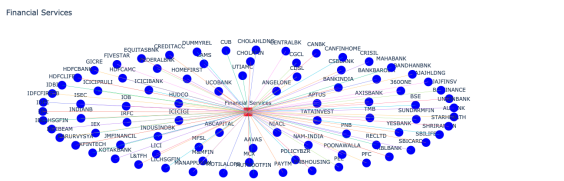

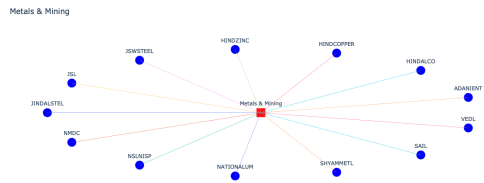

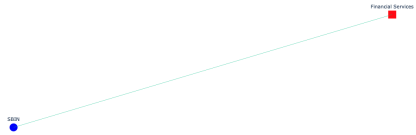
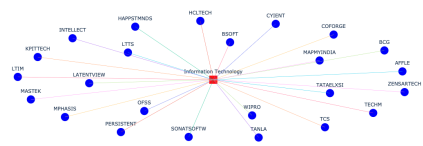

## CPU usage comparison

### Elastic Search

### PostgreSQL

### Neo4J

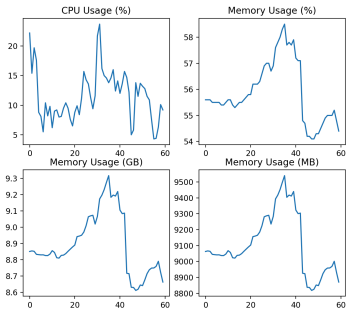
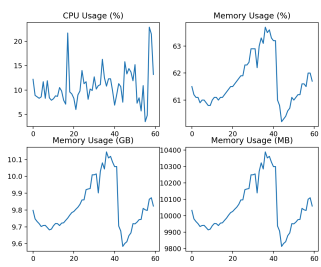
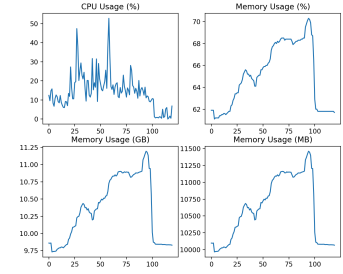

## Configuration

| Item | Value |
| --- | --- |
| OS Name | Microsoft Windows 11 Pro |
| Version | 10.0.22621 Build 22621 |
| Other OS Description | Not Available |
| OS Manufacturer | Microsoft Corporation |
| System Name | |
| System Manufacturer | Microsoft Corporation |
| System Model | Surface Book 2 |
| System Type | x64-based PC |
| System SKU | Surface_Book_1832 |
| Processor | Intel(R) Core(TM) i7-8650U CPU @ 1.90GHz, 2112 Mhz, 4 Core(s), 8 Logical Processor(s) |
| BIOS Version/Date | Microsoft Corporation 394.651.768, 11-Apr-15 |
| SMBIOS Version | 3.3 |
| Embedded Controller Version | 255.255 |
| BIOS Mode | UEFI |
| BaseBoard Manufacturer | Microsoft Corporation |
| BaseBoard Product | Surface Book 2 |
| BaseBoard Version | Not Available |
| Platform Role | Slate |
| Secure Boot State | On |
| PCR7 Configuration | Elevation Required to View |
| Windows Directory | C:\WINDOWS |
| System Directory | C:\WINDOWS\system32 |
| Boot Device | \Device\HarddiskVolume1 |
| Locale | United States |
| Hardware Abstraction Layer | Version = "10.0.22621.1413" |
| User Name | KARTHIK-SURFACE\Karthik |
| Time Zone | Malay Peninsula Standard Time |
| Installed Physical Memory (RAM) | 16.0 GB |
| Total Physical Memory | 15.9 GB |
| Available Physical Memory | 5.56 GB |
| Total Virtual Memory | 29.6 GB |
| Available Virtual Memory | 14.8 GB |
| Page File Space | 13.7 GB |